# Light with a self-torque: extreme-ultraviolet beams with time-varying orbital angular momentum

Laura Rego[1,*], Kevin M. Dorney[2,*], Nathan J. Brooks[2], Quynh Nguyen[2], Chen-Ting Liao[2], Julio San Román[1], David E. Couch[2], Allison Liu[2], Emilio Pisanty[3], Maciej Lewenstein[3,4], Luis Plaja[1], Henry C. Kapteyn[2], Margaret M. Murnane[2], Carlos Hernández-García[1]

[1]*Grupo de Investigación en Aplicaciones del Láser y Fotónica, Departamento de Física Aplicada, University of Salamanca, Salamanca E-37008, Spain*

[2]*JILA - Department of Physics, University of Colorado and NIST, Boulder, Colorado 80309, USA*

[3]*ICFO, Institut de Ciencies Fotoniques, The Barcelona Institute of Science and Technology, Av. Carl Friedrich Gauss 3, 08860 Castelldefels (Barcelona), Spain*

[4]*ICREA, Pg. Lluís Companys 23, 08010 Barcelona, Spain*

laura.rego@usal.es; kevin.dorney@colorado.edu

[*]*These authors contributed equally to this work*

Twisted light fields carrying orbital angular momentum (OAM) provide powerful capabilities for applications in optical communications, microscopy, quantum optics and microparticle rotation. Here we introduce and experimentally validate a new class of light beams, whose unique property is associated with a temporal OAM variation along a pulse: the self-torque of light. Self-torque is a phenomenon that can arise from matter-field interactions in electrodynamics and general relativity, but to date, there has been no optical analog. In particular, the self-torque of light is an inherent property, which is distinguished from the mechanical torque exerted by OAM beams when interacting with physical systems. We demonstrate that self-torqued beams in the extreme-ultraviolet (EUV) naturally arise as a necessary consequence of angular momentum conservation in non-perturbative high-order harmonic generation when driven by time-delayed pulses with different OAM. In addition, the time-dependent OAM naturally induces an azimuthal frequency chirp, which provides a signature for monitoring the self-torque of high-harmonic EUV beams. Such self-torqued EUV beams can serve as unique tools for imaging magnetic and topological excitations, for launching selective excitation of quantum matter, and for manipulating molecules and nanostructures on unprecedented time and length scales.

**Main**

Structured light is critical for a host of applications in imaging and spectroscopy, as well as enhancing our ability to optically manipulate macro- to nano-scale objects such as particles, molecules, atoms and electrons. Particularly in recent years, the novel phase and intensity properties of structured light beams achieved by exploiting the angular momentum of light have garnered renewed interest in optical manipulation and control[1]. One of the most relevant structured light beams are those carrying orbital angular momentum (OAM), also known as vortex beams, introduced by Allen et al. in 1992[2]. The OAM of light manifests from a spatially dependent wavefront rotation of the light beam, which is characterized by the phase winding number, or topological charge, $\ell$. OAM beams have been harnessed for applications in diverse fields[3] such as laser communication[4,5], phase-contrast[6,7] and superresolution microscopy[8], kinematic micromanipulation[9], quantum information[10], and lithography[10]. Spurred by these exciting technologies, a paralleled interest in the ability to control and manipulate the OAM of ultrafast light pulses has also emerged, resulting in numerous techniques that can imprint OAM directly onto an arbitrary waveform. Diffractive optics (e.g., q- and spiral-phase plates)[11,12] and holographic techniques[13] can impart OAM onto waves from radio, to optical, and even x-ray[14] frequencies, while recent advances in high-harmonic generation (HHG) have produced attosecond extreme ultraviolet (EUV) pulses with designer OAM[15-24].

One of the most exciting capabilities enabled by OAM beams is their ability to exert photomechanical torques[2,25,26]. Whereas the linear momentum of light can be employed to control and manipulate microscopic objects via the gradient/scattering force associated with its intensity profile, optically-induced torque manifests from angular momentum transfer between an object and a light field. This enables fundamental capabilities in advanced classical and quantum optical control and manipulation techniques, such as optical tweezers, lattices, and centrifuges[9,27-30], allowing for the realization of molecular and micromechanical rotors, single particle trafficking, and fundamental studies of atomic motion in liquids and Bose-Einstein condensates[31,32].

In this work, we theoretically predict and experimentally validate light beams that carry time-dependent OAM, thus, presenting a self-torque. This new property of structured light, the self-torque, $\hbar\xi$, is defined as $\hbar\xi = \hbar d\ell(t)/dt$, where $\hbar\ell(t)$ is the time-dependent OAM content of the light pulse. The term self-torque follows an analogy with other physical systems that possess a self-induced time-variation of the angular momentum—which, for example, results from the radiation reaction of charged particles[33] or from gravitational self-fields[34]. Although OAM is well understood as a spatial property of light beams, to date, temporarily-dependent OAM light pulses have not been proposed—or observed. We demonstrate that the self-torque arises as a necessary consequence of angular momentum conservation during the extremely non-linear, non-perturbative optical process of high-order harmonic generation (HHG). In HHG, the interaction of an intense field with an atom or molecule leads to the ionization of an electronic wavepacket, which acquires energy from the laser field before being driven back to its parent ion, and emitting a high-frequency photon upon recollision[35,36]. The emitted harmonic radiation can extend from the EUV to the soft x-ray regime if the emissions from many atoms add together in phase[37-40]. The resulting comb of fully coherent harmonics of the driving field in turn yields trains of phase-locked attosecond pulses[41,42].

Self-torqued light beams naturally emerge when HHG is driven by two time-delayed infrared (IR) pulses that differ by one unit of OAM (see Fig. 1). The dynamical process of HHG makes it possible to imprint a "continuous" time-varying OAM, where all OAM components are present—thus creating self-torqued EUV beams. Intuitively, these exotic pulses can be understood as being composed of time-ordered photons carrying consecutively increasing OAM.

The self-torque of light translates to an azimuthal frequency chirp (i.e. a spectral shift along the azimuthal coordinate) on the radiation emission—and vice versa. This allows us to quantify the self-torque by an experimental measurement of the azimuthal frequency chirp. In addition, the degree of self-torque of EUV harmonic beams can be precisely controlled through the time delay and pulse duration of the driving, IR laser pulses. Our work not only presents and confirms an inherently new property of light beams, but also opens up a route for the investigation of systems with time-varying OAM that spontaneously appear in nature[43], as macroscopic dynamical vortices or—thanks to the high frequency of the beams—microscopic ultrafast systems. For example, since short-wavelength light can capture the fastest dynamics in materials[44,45], self-torqued EUV beams serve as unique tools for imaging magnetic and topological excitations, for launching selective and chiral excitation of quantum matter[46], imprinting OAM centrifuges[28], switching superpositions of adiabatic charge migration in aromatic or biological molecules[47,48], or for manipulating the OAM dichroism of nanostructures[49] on attosecond timescales.

**Theory underlying the self-torque of light**

In order to create light beams with self-torque, we drive the HHG process with two linearly polarized IR pulses exhibiting the same frequency content (centered at $\omega_0 = 2\pi c/\lambda_0$), but with different OAM, $\ell_1$ and $\ell_2$, where $|\ell_1 - \ell_2| = 1$. The two laser pulses are separated by a variable time delay, $t_d$, which is on the order of the individual pulse widths (see Methods and Supplementary Information, SI), as shown in Figure 1. These two collinear IR vortex beams are then focused into an atomic gas target, such that the transverse intensity distribution of the two drivers exhibits maximum overlap. We model the HHG process using full quantum simulations in the strong-field approximation (SFA) that include propagation via the electromagnetic field propagator[50], a method that was used in several previous calculations of HHG involving structured pulses[16,19,20,24,40,51,52]. We consider the driving vortex pulses with $\ell_1 = 1$ and $\ell_2 = 2$, described by a $\sin^2$ envelope with $\tau = 10$ fs full-width at half-maximum (FWHM) in intensity, centered at $\lambda_0 = 800$ nm, and delayed by $t_d = \tau = 10$ fs (see Methods for further details).–The inset in Figure 1 shows the time-dependent OAM of the 17$^{th}$ harmonic obtained from our simulations (color background). In order to extract the temporal variation of the OAM, we first select the HHG spectrum in the frequency range $(q-1)\omega_0$ to $(q+1)\omega_0$ (where $q$ is the harmonic order to explore, being $q$=17 in Figure 1), and then we perform a Fourier transform along the azimuthal coordinate[19] at each time instant during the HHG process. Remarkably, the temporal variation of the OAM is monotonic and "continuous", spanning over an entire octave of consecutive topological charges—i.e., it includes all OAM components from $q\ell_1$=17 to $q\ell_2$=34.

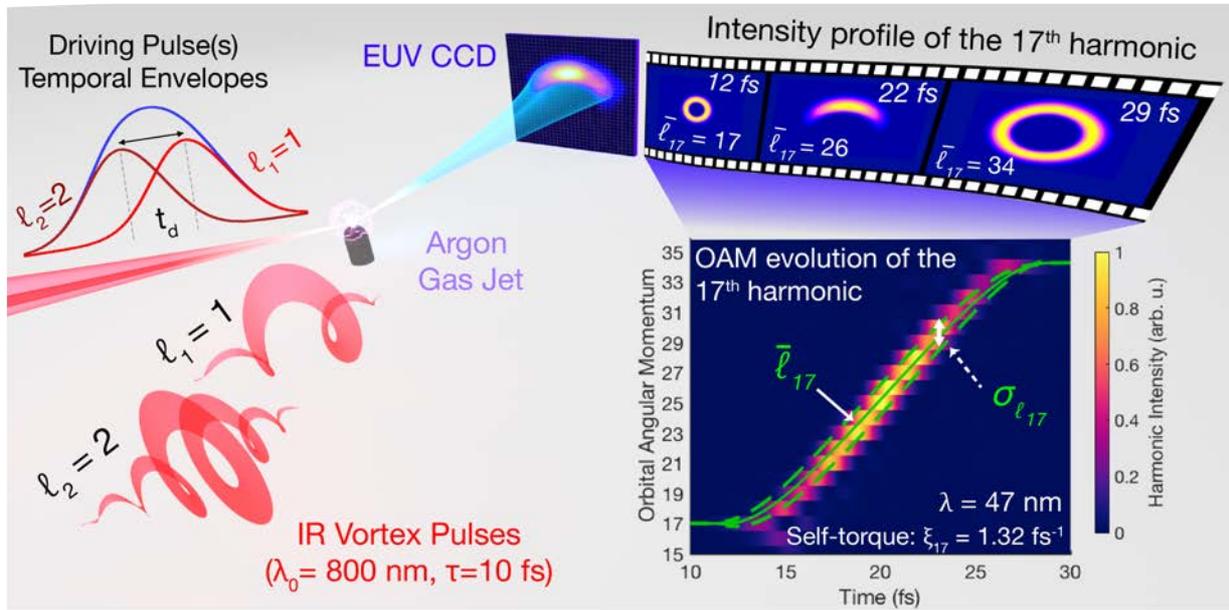

**Figure 1 | Generation of EUV Beams with Self-Torque.** Scheme for the creation of EUV beams with self-torque. Two time-delayed, collinear IR pulses with the same wavelength ($\lambda_1 = \lambda_2 = 800$ nm), but different OAM ($\ell_1 = 1$ and $\ell_2 = 2$), are focused into an argon gas target. The temporal envelopes of each pulse (red), with duration $\tau$ and delayed by $t_d$, and their superposition (blue), are shown on the left. The highly nonlinear process of HHG up-converts the IR pulses into the EUV regime, where harmonic beams with self-torque (i.e., a smooth and continuous time-dependent OAM) are created. The detector shows the spatial profile of the complete, time-integrated HHG beam from our full quantum simulations, while the film strip shows, schematically, the evolution of the intensity profile of the 17$^{th}$ harmonic (47 nm) as a function of time during the emission process. The bottom-right inset represents the temporal evolution of the OAM of the 17th harmonic, for two driving pulses with the same duration, $\tau = 10$ fs, and a time delay of $t_d = \tau$. The color background shows the results from the full quantum simulations, whereas the mean OAM, $\bar{\ell}_{17}$ (solid green), and the width of the OAM distribution, $\sigma_{\ell_{17}}$ (distance between dashed-green lines), are obtained from Eqs. (2) and (3). The self-torque associated to this pulse, $\xi$=1.32 fs$^{-1}$, is obtained from the slope of the time-dependent OAM.

The nature of self-torqued beams can be understood through a simple theoretical analysis. Previous works in OAM-HHG have demonstrated that an IR vortex beam can be coherently converted into high-frequency vortex beams[15-24]. When HHG is driven by a single, linearly polarized, IR vortex beam with integer topological charge, $\ell_1$, the OAM of the $q^{th}$-order harmonic follows a simple scaling rule, $\ell_q = q\ell_1$[16,17]. This scaling reflects the nature of OAM conservation in HHG, where $q$ IR-photons combine to produce the $q^{th}$-order harmonic. If HHG is driven by the combination of two collinear and temporally overlapped IR vortices with different OAM, $\ell_1$ and $\ell_2$, each harmonic order will span over a wide OAM spectrum, given by $\ell_q = n_1\ell_1 + n_2\ell_2$[19], where $n_1$ and $n_2$ are the number of photons absorbed from each driver ($n_1 + n_2 = q$, whose totally must be odd due to parity restrictions). Each channel, ($n_1$, $n_2$), is weighted according to a binomial distribution, associated with the different combinations of absorbing $n_1$ photons with $\ell_1$ and $n_2$ photons with $\ell_2$. Note that the contribution of the harmonic intrinsic phase to the OAM spectrum, also explored in[19], is a second-order effect, negligible for the results presented here.

In this work, however, a delay is introduced between the two IR vortex pulses. The superposition of the delayed envelopes introduces a temporal dependence in the relative weights of the driving fields—thus introducing time as an additional parameter. To show how this influences the OAM structure of the EUV harmonics, we consider two time-delayed, collinear, linearly-polarized, IR driving pulses with different OAM, $\ell_1$ and $\ell_2$. We denote, in cylindrical coordinates $(\rho, \phi, z)$, the complex amplitudes of the driving fields at the focus position ($z$=0) as $U_1(\rho, \phi, t)$ and $U_2(\rho, \phi, t)$. For simplicity, we consider the field amplitudes at the ring of maximum intensity at the target—where the HHG efficiency is highest—and the resulting field can be written as $U(\phi, t) = U_0(t)\{[1 - \eta(t)]e^{i\ell_1\phi} + \eta(t)e^{i\ell_2\phi}\}$, where $U_0(t) = U_1(t) + U_2(t)$ and $\eta(t) = U_2(t)/U_0(t)$ is the relative amplitude of the second beam. According to the strong-field description of HHG, the amplitude of the $q^{th}$-order harmonic, $A_q(\phi, t)$, scales with that of the driving laser with an exponent $p<q$ ($p \simeq 4$ for our laser parameters[19]), while the $q^{th}$-order harmonic phase is considered to be $q$ times that of the driver (see SI Section S1 for the complete derivation), thus

$$A_q(\phi, t) \propto U_0^p(t) \left[\sum_{r=0}^{p} \binom{p}{r} (1-\bar{\eta}(t))^r e^{ir\ell_1\phi} \bar{\eta}^{(p-r)}(t) e^{i(p-r)\ell_2\phi}\right] e^{i(q-p)[(1-\eta(t))\ell_1+\eta(t)\ell_2]\phi}$$

[1]

where $r$ is an integer and $\bar{\eta}(t)$ is the average of $\eta(t)$ over the half-cycle that contributes to the generation of a particular harmonic. The summation in equation (1) is carried over $p$ different OAM channels, each weighted by a binomial distribution according with the combinatory nature of the HHG up-conversion process. Note that parity conservation in HHG demands that the total number of photons absorbed from each driving field, $n_1+n_2$, must be odd, which in turn implies that in order to generate all intermediate OAM states between $q\ell_1$ and $q\ell_2$, the OAM of the drivers must differ by one unit, i.e. $|\ell_1 - \ell_2|=1$. The mean OAM of the $q^{th}$-order harmonic at any instant of time along the harmonic pulse is given by (see SI)

$$\bar{\ell}_q(t) = q[(1-\bar{\eta}(t))\ell_1 + \bar{\eta}(t)\ell_2] \quad [2]$$

and the width of the OAM distribution is

$$\sigma_{\ell_q} = |\ell_2 - \ell_1| \sqrt{p\bar{\eta}(t)(1-\bar{\eta}(t))} \quad [3]$$

In analogy with mechanical systems, we characterize the time-varying OAM spectrum of the $q^{th}$-order harmonic via the self-torque

$$\xi_q = d\bar{\ell}_q(t)/dt \quad [4]$$

Note that as the OAM of light is defined as $\hbar\ell$, the self-torque is given by $\hbar\xi$. For simplicity we factor out $\hbar$ and denote the self-torque by $\xi$, in units of $fs^{-1}$. In the inset of Figure 1, we show the temporal evolution of the mean OAM of the $17^{th}$ harmonic, $\bar{\ell}_{17}$ (solid-green line), and its OAM width, $\sigma_{\ell_{17}}$ (dashed-green lines). In this case, where $t_d = \tau$, we can approximate the self-torque as constant over the OAM span:

$$\xi_q \sim q(\ell_2 - \ell_1)/t_d, \quad [5]$$

which provides a straightforward route for controlling the self-torque through the OAM of the driving pulses and their temporal properties. The example shown in the inset of Figure 1 corresponds to a self-torque of $\xi_{17} = 1.32 \, fs^{-1}$, which implies an attosecond variation of the OAM. Note that equation (5) is valid only if $t_d \simeq \tau$, and if this condition is relaxed, the self-torque must be calculated from the definition given by equation (4). Actually, $t_d = \tau$ is a particularly interesting case, as it corresponds to the time delay where the weight of all intermediate OAM states is more uniform over the OAM span (see Fig. S1 in the SI for the time-dependent OAM for different time delays, showing a consistently excellent agreement between the full quantum simulations and the OAM content predicted by equations (2) and (3)).

It is important to stress that even though the mean OAM value at each instant of time may be non-integer, the nature of self-torqued beams is different from that of the well-known fractional OAM beams[20,53-55]. In particular, the mere superposition of two time-delayed vortex beams—carrying $\ell_i = q\ell_1$ and $\ell_f = q\ell_2$ units of OAM respectively—does not contain a self-torque. Although it does lead to a temporal variation of *the average OAM* similar as in equation (2), it does not contain physical intermediate OAM states, i.e. photons with OAM other than $\ell_i$ and $\ell_f$. Self-torqued beams, on the other hand, contain all intermediate OAM states time-ordered along the pulse (see Fig. 1). In addition, the width of the instantaneous OAM distribution of self-torque beams [equation (3)] is much narrower than that of the mere superposition of time-delayed OAM beams—which in the case of $\ell_i = q\,\ell_1$ and $\ell_f = q\,\ell_2$ is $\sigma_{\ell_q} = q|\ell_2 - \ell_1|\sqrt{\eta(1-\eta)}$. Also, the relative uncertainty, $\sigma_{\ell_q}/\bar{\ell}_q(t)$, decreases with the harmonic order, approaching a classical deterministic regime (see S1 in the SI). This is a result of the non-perturbative behavior of HHG, that enables the creation of well-defined intermediate OAM states in a self-torqued beam. Further non-trivial distinctions between these two different kinds of beam, including their temporal evolution of phase and intensity profiles, are highlighted in the SI.

It is of paramount relevance to evidence the physical nature of the self-torqued beams by temporally characterizing the intermediate OAM states, $\ell_q(t_k)$, with $q\ell_1 < \ell_q(t_k) < q\ell_2$. Assuming a beam with constant self-toque $\xi_q$, the component of the $q^{th}$-order harmonic carrying an OAM of $\ell_q(t_k)$ will appear at the time $t_k = \frac{\xi_q}{\ell_q(t_k)-q\ell_1}$ after the peak amplitude of the first driving pulse, exhibiting a temporal width of $\Delta t_k = \frac{\sigma_{\ell_q}}{\xi_q} = \tau \frac{\sqrt{p\bar{\eta}(1-\bar{\eta})}}{q} \ll \tau$. Therefore, a self-torqued pulse can be thought as a train of time-ordered, overlapping pulses, each one possessing an intermediate OAM, with temporal durations on the attosecond timescale (i.e., much smaller than the width of the driving pulses). This allows us to stress the difference between self-torqued beams and a train of non-overlapping pulses with different OAM[56]. Finally, in analogy to polarization gating techniques[57], self-torqued EUV beams allow the use of subfemtosecond OAM-gating techniques, providing an unprecedented temporal control over laser-matter interactions involving OAM.

**Azimuthal chirp and experimental confirmation of the self-torque of EUV beams**

A direct consequence of self-torque is the presence of an azimuthal frequency chirp in the light beam. As the phase term associated with a time-dependent OAM is given by $\ell_q(t)\phi$, the instantaneous frequency of the $q^{th}$-order harmonic—given by the temporal variation of the harmonic phase, $\varphi_q(t)$—is shifted by the self-torque as:

$$\omega_q(t,\phi) = \frac{d\varphi_q(t)}{dt} = \omega_q + \frac{d\ell_q(t)}{dt}\phi \approx \omega_q + \xi_q \phi \quad [6]$$

Therefore, the harmonics experience an azimuthal frequency chirp whose slope is the self-torque.

In Figures 2a and 2c, we present the HHG spectrum along the azimuthal coordinate obtained in our full quantum simulations for driving pulses of $\tau = 52$ fs and time delays of $t_d = 50.4$ fs and $t_d = -50.4$ fs, respectively, mimicking the experimental parameters (see Methods). Both spectra reflect the presence of an azimuthal chirp, and thus, a self-torque, whose sign depends on $t_d$. The full quantum simulations are in perfect agreement with the analytical estimation given by equation (6) (grey dashed lines). This result shows that the spectral bandwidth of the harmonics can be precisely controlled via the temporal and OAM properties of the driving pulses. Moreover, it provides a direct, experimentally measurable parameter to extract the self-torque, without measuring the OAM of each harmonic at each instant of time with subfemtosecond resolution, which is currently unfeasible. Of course, this reasoning implies that a beam with azimuthal frequency chirp would also exhibit self-torque. Up to now, however, HHG beams have only been driven either by spatially chirped pulses (such as the so-called "attosecond lighthouse" technique[58,59]), or angularly chirped pulses, which (in theory) yield spatially chirped harmonics[60]. However, to the best of our knowledge, azimuthal chirp—and thus, self-torque—has not been imprinted into EUV harmonics, nor in any other spectral regime.

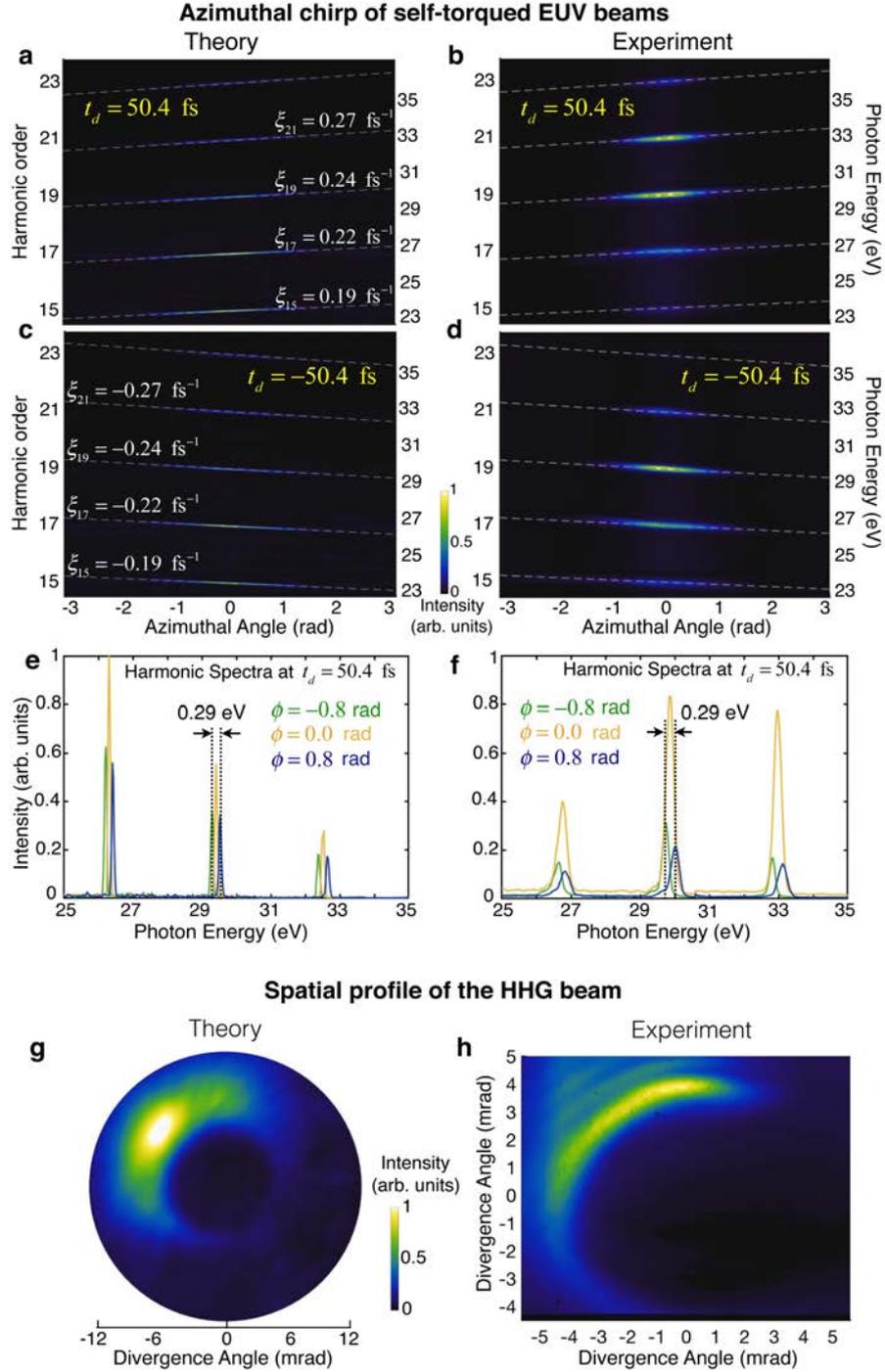

**Figure 2 | Azimuthal Frequency Chirp and Experimental Measurement of the Self-Torque of EUV Beams.** (a-d) Spatial HHG spectrum along the azimuthal coordinate ($\phi$) from quantum simulations (a,c) and experiment (b,d), when the time delay between the driving pulses is (a,b) $t_d$=50.4 fs and (c,d) $t_d$=-50.4 fs. The light self-torque imprints an azimuthal frequency chirp, which is different for each harmonic, as indicated by the grey dashed lines (obtained from equation (6)). Panels (e,f) show the theoretical and experimental harmonic lineouts obtained at $\phi$=-0.8 rad (green), $\phi$=0.0 rad (yellow) and $\phi$=0.8 rad (blue) for $t_d$=50.4 fs. The azimuthal frequency chirp serves as a direct measurement of the self-torque of each harmonic beam. Panels (g, h) present the theoretical and experimental spatial intensities of the HHG beams, after passing through an Al filter, comprising harmonics $q$=13-23.

To experimentally measure the azimuthal frequency chirp of self-torqued EUV beams, we drive the HHG process in argon gas using two collinear, ~52 fs IR vortex beams with topological charges $\ell_1$=1 and $\ell_2$=2. The time delay between the two laser pulses is controlled with a high-resolution linear delay stage, with subfemtosecond repeatability and accuracy. The emitted harmonics are collected by a cylindrical-mirror-flat-grating spectrometer and imaged with an EUV charge-coupled device (CCD) camera (see Methods, SI). Energy is mapped onto one axis of the camera by spectrally dispersing and focusing along the same dimension, while the other axis preserves spatial information. The measured spatial profile of the HHG beam manifests in a "crescent" shape, also found in our simulations (cf. Figures 2g and 2h), which already gives a clear indication of the presence of all intermediate OAM contributions from $q\ell_1$ to $q\ell_2$, and thus, of the creation of self-torqued beams (see Section S2 in the SI). To measure the azimuthal frequency chirp, we exploited the spatiospectral mapping of the spectrometer, by aligning the HHG "crescent" (see SI, Figure S2h) along the spectrometer's dispersion axis, which directly records the azimuthal chirp and yields a tilted harmonic spectrum. Figures 2b and 2d show the experimental spatial profiles of the HHG spectra along the azimuthal coordinate, for time delays of $t_d$ =50.4 and -50.4 fs, respectively. The different slope of the azimuthal chirp, and the excellent agreement with the analytical theory given by equation (6) (grey dashed lines), and also the full quantum simulations shown in Figs, 2a and 2c, confirms the presence of self-torque in the harmonic beams.

In Figure 3, we plot the experimental (solid lines) and theoretical (dashed lines) self-torques obtained for the 17[th] (a), 19[th] (b), 21[st] (c) and 23[rd] (d) harmonics as a function of the time delay between the IR drivers, for the same parameters as in Fig. 2. As the time delay is varied, so too does the degree of azimuthal frequency chirp across the entire harmonic spectrum [according to equations (2) and (6)], verifying the dynamical build-up of OAM in the self-torqued beams. Note that the self-torque is extracted from the measured azimuthal spectral shift (see Figure 2f) and the azimuthal extent of the HHG beam (see SI for details), using equation (6). The excellent agreement, and most importantly, the overall trend, unequivocally demonstrates the presence of a temporally evolving OAM content and, thus, a self-torque, in all the EUV harmonics generated.

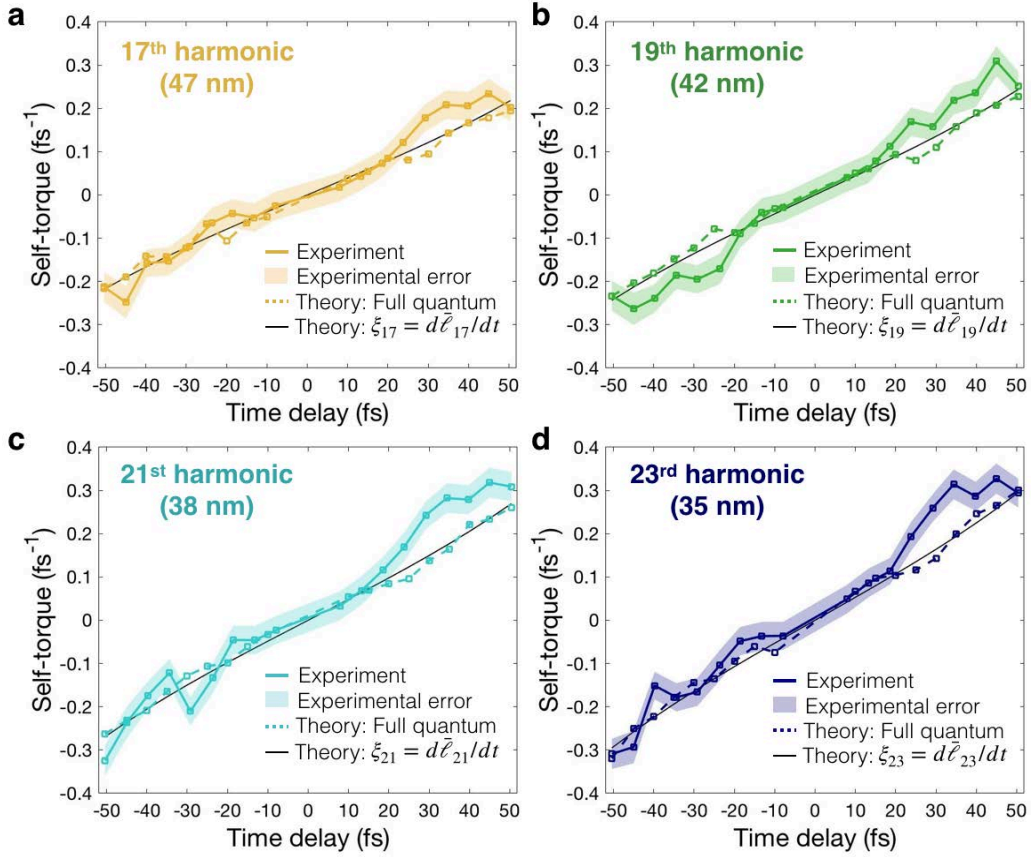

**Figure 3 | Experimental Confirmation of the Self-Torque of Light in EUV Beams.** Self-torques obtained for the 17$^{th}$ (a), 19$^{th}$ (b), 21$^{st}$ (c) and 23$^{rd}$ (d) harmonics as a function of the time delay between the IR drivers. The experimental data is shown in color solid lines the results from full quantum simulations in dashed lines, and the analytical estimation given by equation (2) in solid black lines. The shaded regions depict the experimental uncertainty in the retrieved self-torque for each harmonic order, which themselves comprise the standard "one sigma" deviation of the measured self-torque (i.e., 68% of the measured self-torque values will fall within this uncertainty range).

**Self-torque versus time duration and EUV supercontinuum generation**

Attosecond EUV beams with self-torque can be generated and controlled via the properties of the driving IR vortex beams, with optimal self-torque produced when the laser pulse separation is equal to their duration (i.e., $t_d = \tau$), where all intermediate OAM contributions appear with a similar weight (see Fig. S1 in the SI). To illustrate this concept, Figure 4a shows the simulated self-torque obtained for different IR driving pulse durations.

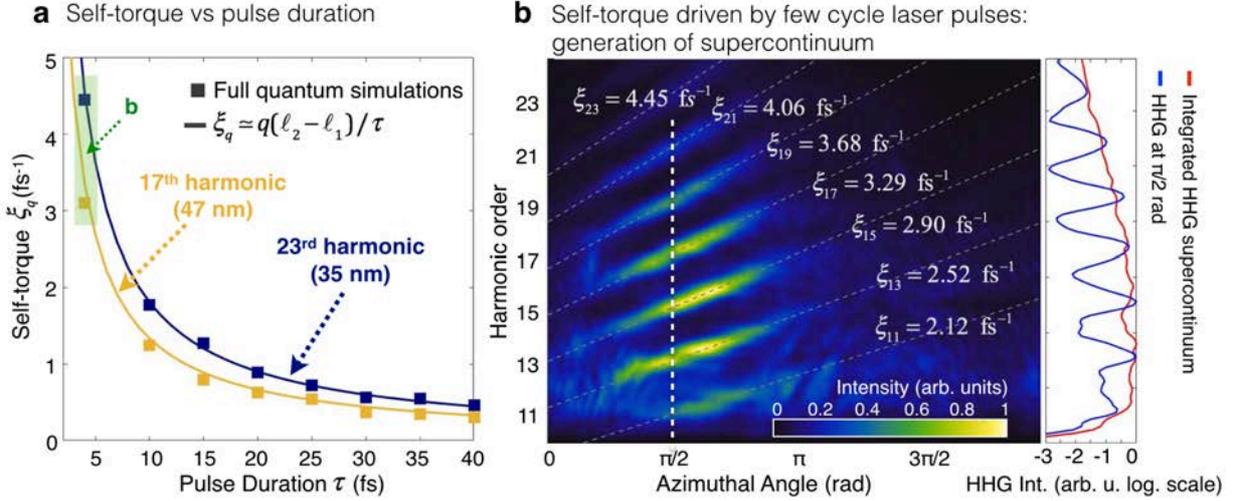

**Figure 4 | Manifestation of Self-Torque for EUV Supercontinuum Generation.** (a) Self-torque versus the pulse duration for the 17$^{th}$ (47 nm) and 23$^{rd}$ (35 nm) harmonics, where $\tau = t_d$. Solid lines are calculated from equation (2), and the squares correspond to results from our full quantum simulations. (b) Spatiospectral HHG distribution when driven by two 800 nm, 4 fs pulses with $\ell_1=1$ and $\ell_2=2$, delayed by 4 fs. The optical self-torque imprints an azimuthal frequency chirp, which is different for each harmonic order, as indicated by the grey dashed lines (obtained from equation (5) and (6)). The right panel shows the HHG yield at π/2 rad (blue line) and the spatially integrated supercontinuum (red line).

In particular, if driven by few-cycle pulses, the self-torque—and thus the azimuthal chirp—is high, with large amounts of OAM building up on an attosecond timescale (Figure 4b, where $\tau = 4$ fs). If the torque is high enough, the harmonic frequency comb sweeps along the azimuth, encapsulating all the intermediate frequencies between the teeth of the harmonic comb. Thus, the frequency chirp of time-dependent OAM beams is not only useful to measure the self-torque, but it also represents an approach to obtain an EUV supercontinuum, as shown in the right inset of Figure 4b. This allows for the creation of a very precise, azimuthally-tunable frequency comb in the EUV, and a supercontinuum spectrum that is complementary, yet distinct, from other approaches[61-63].

**Conclusions**

We have demonstrated that light beams with time-dependent OAM can be created, thus carrying optical self-torque. This property spans the applications of structured light beams[1] by adding a new degree of freedom, the self-torque, and thus introducing a new route to control light-matter interactions. In particular, ultrafast, short wavelength, high harmonic beams with self-torque can be naturally produced by taking advantage of the conservation laws inherent to extreme non-linear optics. This new capability can deliver optical torque on the natural time and length scales of charge and spin ordering (e.g., femtosecond and nanometer), and can lead to new methods for capturing magnetic and topological excitations, for launching selective excitation of quantum matter, or for manipulating molecules and nanostructures. Finally, the self-torque of light imprints an azimuthal frequency chirp which allows a way to experimentally measure and control it. Moreover, if the self-torque is high enough, the harmonic frequency comb sweeps smoothly along the azimuth, and if integrated, a high-frequency supercontinuum is obtained, thus presenting exciting perspectives in EUV and ultrafast spectroscopies of angular momentum dynamics.

**Methods**

**Full quantum SAM-OAM HHG simulations including propagation.** In order to calculate the HHG driven by two time-delayed OAM pulses, we employ a theoretical method that computes both the full quantum single-atom HHG response and subsequent propagation[50]. The propagation is based on the electromagnetic field propagator, in which we discretize the target (gas jet) into elementary radiators[50]. The dipole acceleration of each elementary source is computed using the full quantum SFA, instead of solving directly the time dependent Schrödinger equation, yielding a performance gain in computational time when computing HHG over the entire target[50]. We assume that the harmonic radiation propagates with the vacuum phase velocity, which is a reasonable assumption for high-order harmonics. Propagation effects in the fundamental field, such as the production of free charges, the refractive index of the neutrals, the group velocity walk-off, as well as absorption in the propagation of the harmonics, are taken into account. Note that although we account for the time-dependent nonlinear phase shifts in the driving fields, nonlinear spatial effects are not taken into account. We consider two vortex beams with $\ell_1 = 1$ and $\ell_2 = 2$, whose spatial structure is represented by a Laguerre-Gaussian beam (see equation (S13) in the SI). The laser pulses are modeled with a $\sin^2$ envelope whose FWHM in intensity is $\tau$, and centered at 800 nm in wavelength. The amplitudes of the two fields are chosen to obtain the same peak intensity ($1 \times 10^{14}$ W/cm$^2$) at focus for each driver at the radii of maximum superposition (i.e., the brightest intensity rings overlap spatially). The driving beam waists are chosen to overlap at the focal plane (being $w_1 = 30.0$ μm for $\ell_1$, and $w_2 = w_1/\sqrt{2} = 21.4$ μm for $\ell_2$) where a 10-μm-wide Ar gas jet flows along the perpendicular direction to the beam propagation, with a peak pressure of 667 Pa (5 torr). The low thickness of the gas jet is due to computational time limitations; however, based on our previous results of OAM-HHG[51], we do not foresee any fundamental deviation when considering thicker gas jets closer to the experimental jet employed in this work (a diameter of 150 μm).

**Experimental generation of self-torqued OAM high-harmonic vortices.** The generation of self-torqued high-harmonics is achieved by impinging a pair of collinear, linearly polarized, non-degenerate IR-vortex beams (with topological charges of $\boldsymbol{\ell_1 = 1, \ell_2 = 2}$) onto a supersonic expansion of argon gas. The IR vortex beams are derived from a high-power, ultrafast regenerative amplifier (790 nm, 40 fs, 9 mJ, 1 kHz, KMLabs Wyvern HE®). The near full output of the amplifier is sent into a frequency-degenerate Mach-Zehnder-type interferometer, which separates and later combines the two driving pulses to form the dual-vortex IR driver. In each spatially separated arm of the interferometer, a combination of half-waveplates, faceted spiral phase plates (16-steps per phase ramp, HoloOr), and independent focusing lenses result in each beam possessing linear polarization, non-degenerate topological charges, and the same waist size at focus. Independent irises in each beam path allow for fine tuning of the transverse mode size at focus and are utilized to match the size of the maximum intensity ring for each driver. A high-precision, high-accuracy, and high-repeatability delay stage (Newport, XMS-160S) is used to control the timing between the two driving pulses, with attosecond precision. The pulses are recombined at the output of the interferometer using a low-dispersion beamsplitter, and then sent into the supersonic expansion of argon gas in a vacuum chamber. We take extreme care to ensure that the two arms experience similar dispersion by utilizing the same thickness and design of optics in each arm of the interferometer, which minimizes effects from carrier-to-envelope phase variation in the separate beam paths. Self-torqued high harmonics are generated via the HHG up-conversion process, then dispersed in 1D via a cylindrical-mirror-flat-grating EUV spectrometer and finally collected by a CCD camera (Andor Newton 940). A 200-nm-thick aluminum filter blocks the residual driving light before entering the spectrometer—while passing harmonics over its transmission range, ~17-72 eV—and all harmonic spectra are corrected for the transmission of the EUV beamline. The driving laser modes themselves, both individually and combined, are characterized by a modified Gerchberg-Saxton phase retrieval algorithm, which solves for the phase of a propagating light beam and allows extraction of the OAM content of the IR vortices (see SI and Supplemental Movie S2), thus ensuring high-quality driving pulses. Note that this modified Gerchberg-Saxton method acquires and retrieves OAM content much faster than our previous characterization method using ptychography[64], but it is limited to non-multiplexed (i.e., single-color) beams. We also note that the use of a frequency-degenerate Mach-Zehnder interferometer results in a 50% intensity loss of each driver when combined at the interferometer's exit; however, this configuration proved ideal to minimize pulse dispersion, while also allowing for independent control of the polarization and topological charge of the driving beams.

**EUV measurements of self-torque.** A measurement of the self-torque of our EUV beams in the time-domain would require both a high-resolution spatial measurement of the harmonic phase, coupled with sub-femtosecond time resolution, which is currently unfeasible. In order to circumvent this limitation, we rely upon a frequency domain measurement of the azimuthal chirp and then exploit the fundamental phase relationship between a temporal phase and instantaneous frequency to extract the self-torque of our experimentally generated high harmonics. In order to measure the azimuthal frequency chirp, we harness a unique property of the physics of superposed OAM beams and the optics of our spectrometer system. When two OAM beams with neighboring, but low, topological charges are superposed, a "crescent" shaped intensity distribution is obtained, where the region of maximum intensity is located at an angle that depends upon the time delay between the two beams. As the time delay is scanned, the intensity crescent rotates about the common origin of the two IR vortex drivers (see Section S5 in the SI). By carefully adjusting the time delay, we can align the center-of-mass of the intensity of the driving crescent with the dispersion axis of the EUV spectrometer, which in turn aligns the high-harmonic intensity crescent (since to first order, the harmonics mimic the intensity distribution of the driver) to the spectrometer as well. Briefly (see SI for more details), the HHG intensity crescent is aligned such that it is directed along the vertical lab-frame coordinate, and when hitting the cylindrical mirror (with its curved axis orthogonal to the propagation direction of the HHG intensity crescent), the HHG beam is collapsed spatially, preserving the divergence and azimuthal angular range of the unfocused HHG beam. The focusing HHG beam then impinges on the grating—which has grooves oriented parallel to the curved axis of the cylindrical mirror—, which disperses the harmonic spectrum along the transverse coordinate of the HHG beam. In such a configuration, the spectrometer naturally disperses the harmonics such that the azimuthal frequency chirp is mapped into the dispersed spectra, yielding spectrally tilted high harmonics. The self-torque for each harmonic order, as a function of time delay, is then extracted from the spectral tilt of the harmonics and a concurrent measurement of the azimuthal extent of the undispersed harmonic beam (see SI Section S6), all while keeping the generation chamber and driving beamline undisturbed. Extreme care is taken to ensure that the harmonic beam remains fixed to the dispersion axis of the spectrometer, while potential artifacts from improper imaging conditions and overdriving of the HHG process are minimized, if not non-existent (see SI, Section S7). Both long- and short-term drifts in beam pointing are minimized by enclosing the entire beamline from external air currents and vibrations, while fluctuations and drifts in pointing are actively stabilized via a home-built pointing stabilization system (see SI, Section S3).

**Acknowledgments**
C.H.-G., J.S.R. and L.P. acknowledge support from Junta de Castilla y León (SA046U16) and Ministerio de Economía y Competitividad (FIS2016-75652-P) and Ministerio de Ciencia, Innovación y Universidades (EQC2018-0041 17-P). C.H.-G. acknowledges support from a 2017 Leonardo Grant for Researchers and Cultural Creators, BBVA Foundation. L.R. acknowledges support from Ministerio de Educación, Cultura y Deporte (FPU16/02591). H. K. and M. M. graciously acknowledge support from the Department of Energy BES Award No. DE-FG02-99ER14982 for the experimental implementation, as well as a MURI grant from the Air Force Office of Scientific Research under Award No. FA9550-16-1-0121 for the theory. N. J. B., Q. L. N., and D. C. acknowledge support from National Science Foundation Graduate Research Fellowships (Grant No. DGE-1144083). E.P. acknowledges Cellex-ICFO-MPQ fellowship funding; E.P. and M.L. acknowledge the Spanish Ministry MINECO (National Plan 15 Grant: FISICATEAMO No. FIS2016-79508-P, SEVERO OCHOA No. SEV-2015-0522, FPI), European Social Fund, Fundació Cellex, Generalitat de Catalunya (AGAUR Grant No. 2017 SGR 1341 and CERCA/Program), ERC AdG OSYRIS, EU FETPRO QUIC, and the National

Science Centre, Poland-Symfonia Grant No. 2016/20/W/ST4/00314. We thankfully acknowledge the computer resources at MareNostrum and the technical support provided by Barcelona Supercomputing Center (RES-AECT-2014-2-0085). This research made use of the high-performance computing resources of the Castilla y León Supercomputing Center (SCAYLE, https://www.scayle.es/), financed by the European Regional Development Fund (ERDF).

**Author contributions**

L.R., J.S.R, L.P. and C.H.-G. conceived the idea of self-torqued beams. K.M.D., L.R., H.C.K., M.M.M. and C.H.-G. designed the experiment. K.M.D., N.J.B., Q.N., C.-T.L., D.C., A.L. conducted the experiment. K.M.D. analyzed the experimental data. L.R., J.S.R., L.P. and C.H.-G. performed the theoretical simulations and analyzed the resulting data. C.H.-G., L.P., M.M.M., and H.C.K. supervised the theoretical simulations, experimental work and developed the facilities and measurement capabilities. L.R., K.M.D., J.S.R., M.M.M., L.P. and C.H.-G. wrote and prepared the manuscript. All authors provided constructive improvements and feedback to this work.

**Financial interests**

M. M. M. and H. C. K. have a financial interest in KMLabs. The other authors declare no competing financial interests.

**Corresponding Authors**

Please address all correspondence to Laura Rego and Kevin M. Dorney.